**Development of an Ion-Sensor using Fluorescence Resonance Energy Transfer**

Dibyendu Dey, Jaba Saha, D. Bhattacharjee, Syed Arshad Hussain*

Department of Physics, Tripura University, Suryamaninagar – 799022, Tripura, India

* Corresponding author

Email: sa_h153@hotmail.com, sah.phy@tripurauniv.in

Ph: +919862804849 (M), +91381 2375317 (O)

Fax: +913812374802

ABSTRACT

A method is presented for the sensing of ions by determining the concentration of corresponding salts (KCl, NaCl, $MgCl_2$, $CaCl_2$, $FeCl_3$, $FeSO_4$, $AlCl_3$) in water, based on Fluorescence resonance energy transfer (FRET) process. The principle of the proposed sensor is based on the change of FRET efficiency between two laser dyes Acriflavine and Rhodamine B in presence of different ions ($K^+$, $Na^+$, $Mg^{2+}$, $Ca^{2+}$, $Fe^{2+}$, $Fe^{3+}$, $Al^{3+}$). Nanodimensional clay platelet laponite was used to enhance the efficiency of the sensor.

# 1. Introduction

The development of ion sensor technology has emerged as a dynamic approach for identifying and quantitating specific analytes of environment. Now a day the growing need for multicomponent analyses and shorter sample preparation methods, new sensing techniques with decreasing costs are very important. There are several sensing techniques which include ion selective or gas sensitive electrodes, thermistors, chemically and biologically modified metal or semiconductors [1-3]. Chalcogemide glass sensors are used for the detection of millimole levels of copper, iron, chromium, lead, cadmium and mercury in natural waste waters [4]. The method is also used for the detection of heavy metal ions in solutions [4]. But the sensing of ions present in micromole amount could be much interesting and useful. It will be very much interesting to sense $K^+$ and $Na^+$ ions in blood samples. In blood the $K^+$ and $Na^+$ concentrations are 4.5 and 120 mM respectively [5]. Increase in concentration of these ions in blood can cause serious health problems. Another type of sensor is fluorescent ion sensors. There are few reports where the detection of heavy metal ions such as $Hg^{2+}$, $Pb^{2+}$ and $Cd^{2+}$ have been done by fluorescent ion sensors with high sensitivity and simplicity [6-9]. But in this method the change in fluorescent intensity could be perturbed by environmental factors [10, 11]. It is interesting to mention in this context that the introduction of ratiometric sensors can minimize this environmental perturbation, because it measures the ratio of two emissions in different environment. The design of ratiometric sensors can be done by two method (i) ICT (intermolecular charge transfer) and (ii) FRET (Fluorescence Resonance Energy Transfer). For many ICT based ion sensors it is difficult to determine the ratio between two relatively broad signal emissions. Recently FRET based sensing has become most effective method for the detection of ions in environment. FRET based sensors have been widely used in metal ion detection [12, 13], FRET sensing of the fluorophores [14-18], Silica [19, 20], and polymer particles [21-23]. In one of our recent paper we have used FRET for the sensing of permanent hard water components in water for a concentration range of 0.03 mg/ml to 0.2 mg/ml [24].

Here in the process of designing ion sensor based on FRET process, we have used two dyes Acriflavine (Acf) and Rhodamine B (RhB) as energy donor and acceptor. In principle both the dyes are suitable for fluorescence resonance energy transfer. Both the dyes are highly fluorescent and the fluorescence spectrum of Acf sufficiently overlaps with the absorption spectrum of RhB. P. D. Sahare et al [25] reported the fluorescence resonance energy transfer in binary solution mixture of these two dyes. Recently we have demonstrated a pH sensor [26], DNA sensor [35] and hard water sensor [24] based on the FRET between Acf and RhB.

In the present communication we tried to investigate the effect of $K^+$, $Na^+$, $Mg^{2+}$, $Ca^{2+}$, $Fe^{2+}$, $Fe^{3+}$, $Al^{3+}$ on the FRET efficiency between two fluorophores, Acf and RhB in presence and absence of nanoclay sheet laponite. The energy transfer efficiency has been effected if the distance between the donor – acceptor pair has been altered due to the presence of any external agency or change of the microenvironment. It has been observed that when distance between fluorophores (dyes) is decreased due to adsorption of the dyes on to nanoclay sheet, the FRET efficiency increases [26, 27]. Here we have used nanoclay sheet laponite to enhance the sensitivity of sensing. Our investigation showed that FRET efficiency decreases with increasing salt concentration. It has also been demonstrated that with proper calibration, FRET between Acf and RhB can be used to sense different ions on the basis of their size and charge up to micromole level.

## 2. Materials and methods

2.1. Material

Both the dyes Acf and RhB were purchased from Sigma Chemical Co. USA and used as received. Ultrapure Milli-Q water (resistivity 18.2 MΩ-cm) was used as solvent. The dyes used in our studies are cationic in nature. The clay mineral used in the present work was Laponite, obtained from Laponite Inorganic, UK and used as received. The size of the clay platelet is less than 0.05 μm and CEC is 0.739 meq/g determined with CsCl [28]. All the salts KCl, NaCl, $MgCl_2$, $CaCl_2$, $FeCl_3$, $FeSO_4$, $AlCl_3$ were purchased from Thermo Fisher Scientific India Pvt. Ltd. and used as received. Dye and salt solutions were prepared in Milli-Q water. For spectroscopic measurement the dye solution concentration was optimized at $10^{-6}$ M. The clay dispersion was prepared by using Milli-Q water and stirred for 24 hours with a magnetic stirrer followed by 30 minutes ultrasonication before use. The concentration of clay was kept fixed at 2 ppm throughout the experiment. To check the effect of clay on the spectral characteristics, the dye solutions (Acf and RhB) were prepared in the clay suspensions (2 ppm). In order to check the effect of salt on spectral characteristics in presence of clay, first of all the salts were added in the clay dispersion individually at different concentration. Then the dyes were added in the salt mixed clay dispersions. In all cases the clay concentration was 2 ppm and the dye concentration was $10^{-6}$ M.

2.2. UV–Vis absorption and fluorescence spectra measurement

UV–Vis absorption and fluorescence spectra of the solutions were recorded by a Perkin Elmer Lambda-25 Spectrophotometer and Perkin Elmer LS-55 Fluorescence Spectrophotometer respectively. For fluorescence measurement the excitation wavelength was 420 nm (close to the absorption maxima of Acf).

## 3. Results and discussions

3.1. FRET between Acf and RhB in aqueous solution and clay dispersion

The absorption and emission maxima of Acf are centered at 449 and 502 nm respectively which is assigned due to the Acf monomers [25]. On the other hand RhB absorption spectrum possess prominent intense 0-0 band at 553 nm along with a weak hump at 520 nm which is assigned due to the 0-1 vibronic transition [29]. The RhB fluorescence spectrum shows prominent band at 571 nm which is assigned due to the RhB monomeric emission [29]. The corresponding absorption and emission spectra of the above results are shown in figure 1 of the supporting information.

Figure 1a shows the fluorescence spectra of pure Acf, RhB and their mixture in water solution in presence and absence of salt (KCl, $MgCl_2$, and $FeCl_3$). All the spectra were measured with excitation wavelength 420 nm (close to absorption maximum of Acf). This excitation wavelength was choosen in order to excite the Acf molecule directly and to avoid the direct excitation of RhB molecule. From the figure it has been observed that the fluorescence intensity of pure Acf (curve 1, of figure 1a) is much higher, on the other hand the fluorescence intensity of pure RhB (curve 2, of figure 1a) is almost negligible. However, the Acf-RhB mixture fluorescence spectrum is (curve 3, of figure 1a) very interesting. Here the Acf emission decreases with respect to pure Acf and on the other hand RhB emission increases with respect to pure RhB (curve 3, of figure 1a). This is mainly due to the transfer of energy from Acf molecule to RhB molecule via fluorescence resonance energy transfer. In order to confirm this, excitation spectra was recorded with monitoring emission wavelength 500 nm (Acf emission maximum) and 571 nm (RhB emission maximum) and observed that both the excitation spectra are very similar to the absorption spectrum of Acf monomer (figure 2 of supporting information). This confirms that the RhB fluorescence is mainly due to the light absorption by Acf and corresponding transfer to RhB monomer. Thus FRET between Acf to RhB has been confirmed. FRET efficiency have been calculated using the following equation [30]

$$E = 1 - \frac{F_{DA}}{F_D}$$

Where $F_{DA}$ is the fluorescence intensity of the donor in the presence of acceptor and $F_D$ is the fluorescence intensity of the donor in the absence of the acceptor.

In order to sense different ions we have introduced different salts in the Acf-RhB mixed aqueous solution and the FRET between Acf to RhB has been measured. The change in FRET efficiency due to the presence of ions / salts has been examined in order to sense the presence of corresponding ions. Fluorescence spectra of Acf-RhB mixture in presence of KCl (curve 4), $MgCl_2$ (curve 5), $FeCl_3$ (curve 6) have also been shown in figure 1a. It has been observed that in all the cases the FRET efficiency decreased. However, the change in FRET efficiency is very small. The FRET efficiency changes from 11.37% (in absence of salt) to 9.2% (in presence of KCl) or 7.4% (in presence of $MgCl_2$) or 5.2% (in presence of $FeCl_3$). The corresponding efficiencies are listed in table 1.

In order to increase the FRET efficiency between Acf to RhB we have introduced nanoclay platelet laponite in the Acf-RhB mixture. Fluorescence spectra of pure Acf (curve 1), RhB (curve 2) and Acf-RhB mixture (curve 3) in presence of laponite are shown in figure 1b. The corresponding FRET efficiencies are also listed in table 1. It has been observed that FRET efficiency increases to 78.17% in presence of clay for Acf-RhB mixture which was 11.37% in absence of clay platelet. Now salts are introduced in the Acf-RhB mixture in presence of clay laponite. Corresponding fluorescence spectra (KCl (curve 4), $MgCl_2$ (curve 5), $FeCl_3$ (curve 6), are also shown in figure 1b. The FRET efficiencies are also listed in table 1.

It is worthwhile to mention in this context that clay particles are negatively charged and have layered structure with a cation exchange capacity [31, 32]. Both the dyes Acf and RhB under investigation are positively charged. Accordingly they are adsorbed onto the clay layers. On the other hand FRET process is very sensitive to distances between the energy donor and acceptor and occurs only when the distance between the D-A pair is of the order of 1-10 nm [33, 34]. Therefore in the present case, clay particles play an important role in determining the concentration of the dyes on their surfaces or to make possible close interaction between energy donor and acceptor in contrast to the pure aqueous solution.

It is important to mention in this context that in one of our earlier works it has been observed that the presence of nanoclay laponite increased the FRET efficiency between N, N'-dioctadecyl thiacyanine perchlorate (NK) and octadecyl rhodamine B chloride (RhB) [27]. Effect of nanoclay laponite on the energy transfer efficiency between Acf and RhB has also been studied [26]. It has been observed that the presence of clay platelet increase the energy transfer efficiency.

From table 1 it has been observed that FRET efficiency increases in presence of nanoclay platelets. Again the presence of salts causes a decrease in FRET efficiencies. Here in presence of clay the change or variation of FRET efficiency due to presence of salts is more compared to that in absence of clay platelet. Therefore incorporation of clay laponite in the present system lower the error level in sensing different ions or increases the ion sensing efficiency/sensetivity.

3.2. Ions with variable size:

In order to have idea about the effect of ion size on FRET efficiency, we have selected three different sets of salts – (i) NaCl and KCl (both are monovalent), (ii) $MgCl_2$ and $CaCl_2$ (both are divalent), (iii) $FeCl_3$ and $AlCl_3$ (both are trivalent), and measured the FRET efficiency between Acf to RhB in presence of these salts. All the experiments were performed in presence of clay.

Figure 2a shows the fluorescence spectra of Acf+RhB (1:1 volume ratio) in presence and absence of NaCl and KCl along with pure Acf and RhB fluorescence spectra. The corresponding energy transfer efficiencies are listed in table 2. It has been observed that the energy transfer efficiencies decrease in presence of different salts. The interesting thing is that the transfer of energy from Acf to RhB is larger in case of NaCl than KCl. It is very much possible due to the fact that the molecular size of $Na^+$ ion is smaller than $K^+$ ion. So the space occupied by the $K^+$ ions on the clay templates is larger than the $Na^+$ ions resulting in a larger intermolecular separation and smaller energy transfer between the dye molecules. The similar kind of study is also done for the divalent and trivalent ions and is shown in figure 2b and c respectively. The corresponding FRET efficiencies are also tabulated in table 2. The trends of energy transfer efficiencies are similar to that for monovalent salts. Here both these two cases it has been observed that energy transfer efficiency is larger for the salts with smaller ion sizes. In later section of this manuscript this has been explained with schematic diagram.

3.4. Ions with variable valency and same size

In order to check the effect of valency of ions on the energy transfer efficiency we have measured the fluorescence spectra of Acf+RhB mixture (1:1 volume ratio) in aqueous clay dispersion in absence and presence of salt $FeSO_4$ (divalent) and $FeCl_3$ (trivalent). Figure 3 shows the corresponding spectra. Energy transfer efficiencies calculated from the spectra of figure 3 are listed in table 3. From figure and calculated values of efficiencies it has been observed that the energy transfer efficiency decreases in presence of both the salts. However, the extent of decrease in energy transfer efficiency is more in presence of $FeCl_3$ compared to that in presence of $FeSO_4$. In this study we have selected the salts in such a way that their molecular size remains same but their charge changes. Here $FeSO_4$ provides a $Fe^{2+}$ ion and $FeCl_3$ provides a $Fe^{3+}$ ion in aqueous medium. So the larger electrostatic repulsion of the $Fe^{3+}$ ion provides a larger intermolecular separation between the Acf and RhB in compared to $Fe^{2+}$ ion on to the clay templates when the ions and dyes were adsorbed on to clay surface. As a result the FRET efficiency decreases more in case of $FeCl_3$. In later section of this manuscript this has also been explained with schematic diagram.

3.5. Effect of variation of salt concentration on FRET efficiency

In the previous sections we have seen that presence of laponite particle increases the FRET efficiency between Acf and RhB, whereas, presence of salt/ions decreases the FRET efficiency. Again the decrease in energy transfer efficiency is proportional to the ion size and valency. Now in order to check the effect of variation of salt concentration on the FRET efficiency, we have measured the fluorescence spectra of Acf and RhB mixture with different salt concentration (10, 100 and 1000 μM) in clay dispersion and the energy transfer efficiency have been calculated. The spectra are available in figure 3 of the supporting information and the corresponding efficiencies are listed in table 4. It has been observed that the FRET efficiency decreases with increasing salt concentration for all the salts. The increase in salt concentration basically increases the amount of cations in the solvent and as a result a larger area of the clay layers is occupied by the salt cations. Accordingly a comparatively smaller amount of dye molecules are attached to the clay templates resulting in a less probability of occurrence of FRET between Acf and RhB.

As a whole our investigations suggest that it is possible to sense the ions by observing the change in FRET efficiency with ion size, valency and varying salt concentration.

3.6. Schematic diagram

A schematic diagram showing the organization of Acf and RhB in absence and presence of clay laponite and salt is shown in figure 4. Normally in absence of clay and salt the distance between Acf and RhB molecules in aqueous solution is larger resulting lower energy transfer efficiency but in presence of clay the dyes are adsorbed by cation exchange reaction on to the clay surface and accordingly the distance between Acf and RhB decreases as shown in the figure 4b. These results can increase the energy transfer efficiency. In presence of both clay and salt, the probability of adsorption of salt cations are larger in compared to the cationic dyes. This is because in the process of dye-clay-salt solution preparation initially the salt was added to the clay dispersion followed by the dye addition. Accordingly, most of the negative charges on the clay surface are neutralized by the salt cations and there exist very few unoccupied negative charges on the clay surface to adsorb the cationic dyes (figure 4c-4f). Accordingly the average distance between Acf and RhB become larger. This results a decrease in energy transfer efficiency. Now as the size of the salt cations increase the space on the clay template decreases farther for the dye molecules to be attached even if the number of salt molecules are same and as a result the FRET efficiency also decreases farther (figure 4c and 4d). On the other hand if the size of the ions are kept constant but the valancy is increased then also there will be a farther decrease of the FRET efficiency for the larger valancy of the salt ions (figure 4e and 4f) because the number of negative ions present on the clay template are neutralized more by the high valency salt molecule than the low valency salt molecule.

3.7. Design of ion sensor

Based on the variation of FRET efficiency or fluorescence intensity, depending on the type of salts we have demonstrated an ion sensor. In the process of ion sensing first of all clay (laponite) dispersion will be prepared using the sample water (in presence of ions) followed by addition of dyes (Acf and RhB). Then the fluorescence spectra of the solution will be measured. By observing the fluorescence intensity or FRET efficiency calculated from the observed fluorescence spectra it would be possible to sense different ions of the sample water.

Figure 5 shows the plot of FRET efficiency as a function of different salts (KCl, NaCl, $MgCl_2$, $CaCl_2$, $FeSO_4$, $FeCl_3$, $AlCl_3$) of concentration 10μM. The data are taken from spectra shown in figures 1(b), 2(a), (b), (c) and 3. From figure 3(a) it has been observed that the FRET efficiency for KCl and NaCl are 56.8% and 62.15% respectively. If the FRET efficiency is observed to be higher than 56.8% but lower than 62.15% then it shows the presence of $Na^+$ ion whereas, if the efficiency is lower 56.8% then it shows the presence of $K^+$ ion. Therefore with proper calibration it is possible to design an ion sensor which can sense ions on the basis of their size (figure 5a-5c) and similar observation can be done for the sensing of ions of different valency but same size (as shown in figure 5b).

It is interesting to mention that in principle this method can be used to sense negative ions also. The presence of negative ions in between the cationic dye pair can decrease their intermolecular separation and the FRET efficiency will increase. Therefore with proper calibration and selecting suitable FRET pair it is possible to sense both negative as well as positive ions using this method. Detailed investigations are required for the sensing of negative ions. Work is going on in our laboratory in this line.

## 4. Conclusion

Fluorescence resonance energy transfer (FRET) between two fluorescent dyes Acriflavine and Rhodamine B were investigated successfully in solution in presence and absence of clay mineral particle laponite. UV-Vis absorption and fluorescence spectroscopy studies reveal that both the dyes present mainly as monomer in solution and there exist sufficient overlap between the fluorescence spectrum of Acf and absorption spectrum of RhB, which is a prerequisite for the FRET to occur from Acf to RhB. Energy transfer occurred from Acf to RhB. The energy transfer efficiency increases in presence of clay laponite in solution. The maximum efficiency was found to be 78.17% for the mixed dye system (50% RhB + 50% Acf) in clay dispersion. In presence of KCl, NaCl, $MgCl_2$, $CaCl_2$, $FeSO_4$, $FeCl_3$, and $AlCl_3$ the FRET efficiency is decreased to 56.8%, 62.15%, 47.4%, 41.1%, 46.8%, 38.5% and 45.6% respectively from 78.17%. With suitable calibration of these results it is possible to design an ion sensor that can sense the presence of different ions in water up to a concentration of 10μM or more.

**Acknowledgements**

The author SAH is grateful to DST and CSIR for financial support to carry out this research work through DST Fast-Track project Ref. No. SE/FTP/PS-54/2007, CSIR project Ref. 03(1146)/09/EMR-II.

**Authors Biography:**

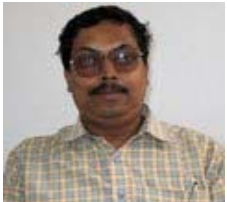

**Prof. D. Bhattacharjee** (M.Sc, Kalyani University & Ph.D, IACS, India) is a Professor in the Department of Physics, Tripura University, India. His major fields of interest are preparation and characterization of ultra thin films by Langmuir-Blodgett & Self-assembled techniques. He visited Finland and Belgium for postdoctoral research. He has undertaken several research projects. He has published more than 66 research papers in different national and international journals and attended several scientific conferences in India and abroad.

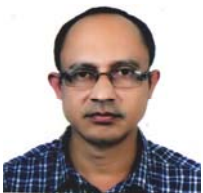

**Dr. S. A. Hussain** (M.Sc. 2001 & Ph.D, 2007, Tripura University, India) is an Assistant Professor in the Department of Physics, Tripura University. His major fields of interest are Thin Films and Nanoscience. He was a Postdoctoral Fellow of K.U. Leuven, Belgium (2007-08). He received Jagdish Chandra Bose Award 2008-2009, TSCST, Govt. of Tripura; Young Scientist Research Award by DAE, Govt of India. He has undertaken several research projects. He has published 56 research papers in international journals and attended several scientific conferences in India and abroad.

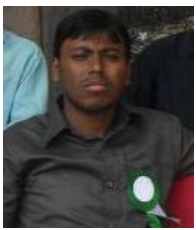

**Mr. Dibyendu Dey** (M.Sc 2009, Tripura University, India) is working as a Research scholar in Department of Physics, Tripura University. His major fields of interest are Fluorescence Resonance Energy Transfer in solution & ultrathin films. He has published 3 research papers in international journals and attended several scientific conferences in India.

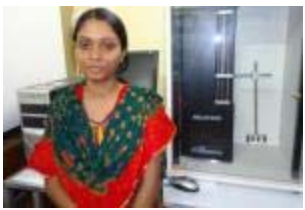

**Miss. Jaba Saha** (M.Sc 2012, Tripura University, India) is working as a Research scholar in Department of Physics, Tripura University. Her major fields of interest are Fluorescence Resonance Energy Transfer in solution & ultrathin films and their sensing applications.

**Table caption**

**Table 1** Values of energy transfer efficiency (E %) for Acf and RhB mixture (1:1 volume ratio) with different salts (KCl, MgCl$_2$, FeCl$_3$) in absence and presence of clay. Dye concentration was 10$^{-6}$M, salt concentration was 10 µM and clay concentration was 2 ppm. The values are calculated from the spectra of figure 1.

**Table 2** Values of energy transfer efficiency (E %) for Acf and RhB mixture (1:1 volume ratio) with different salts (KCl, NaCl, MgCl$_2$, CaCl$_2$, FeCl$_3$, AlCl$_3$) in presence of clay. Dye concentration was 10$^{-6}$M, salt concentration was 10 µM and clay concentration was 2 ppm. The values are calculated from the spectra of figure 2.

**Table 3** Values of energy transfer efficiency (E %) for Acf and RhB mixture (1:1 volume ratio) with different salts (FeSO$_4$, FeCl$_3$) in presence of clay. Dye concentration was 10$^{-6}$M, salt concentration was 10 µM and clay concentration was 2 ppm. The values are calculated from the spectra of figure 3.

**Table 4** Values of energy transfer efficiency (E %) for Acf and RhB mixture (1:1 volume ratio) with different salt concentration in presence of clay. Dye concentration was 10$^{-6}$M and clay concentration was 2 ppm. The values are calculated from the spectra of figure 3 given in the supporting information.

**Figure caption**

**Fig.1. (a)** Fluorescence spectra of pure Acf (1), pure RhB (2) and Acf+RhB (1:1 volume ratio) in water solution (3), with KCl (4), $MgCl_2$ (5), $FeCl_3$ (6). **(b)** Fluorescence spectra of pure Acf with clay (1), pure RhB with clay (2) and Acf+RhB (1:1 volume ratio) in clay suspension (3), with KCl (4), $MgCl_2$ (5), $FeCl_3$ (6). Dye concentration was $10^{-6}$M and clay concentration was 2 ppm and salt concentration was $10^{-5}$ M.

**Fig.2. (a)** Fluorescence spectra of pure Acf with clay (1), pure RhB with clay (2) and Acf+RhB (1:1 volume ratio) in clay suspension (3), with KCl (4), NaCl (5) and **(b)** Fluorescence spectra of pure Acf with clay (1), pure RhB with clay (2) and Acf+RhB (1:1 volume ratio) in clay suspension (3), with $CaCl_2$ (4), $MgCl_2$ (5) **(c)** Fluorescence spectra of pure Acf with clay (1), pure RhB with clay (2) and Acf+RhB (1:1 volume ratio) in clay suspension (3), with $FeCl_3$ (4), $AlCl_3$ (5). Dye concentration was $10^{-6}$M and clay concentration was 2 ppm and salt concentration was $10^{-5}$ M.

**Fig.3.** Fluorescence spectra of pure Acf with clay (1), pure RhB with clay (2) and Acf+RhB (1:1 volume ratio) in clay suspension (3), with $FeCl_3$ (4), $FeSO_4$(5). Dye concentration was $10^{-6}$M and clay concentration was 2 ppm and salt concentration was $10^{-5}$ M.

**Fig. 4.** Schematic representation of FRET between Acf and RhB in presence of (b) clay, (c) clay and salt with larger size ions, (d) clay and salt with smaller size ions, (e) clay and salt with trivalent ions, (f) clay and salt with divalent ions.

**Fig. 5.** FRET efficiency of Acf and RhB mixture for (a) KCl (1), NaCl (2) and without salt (3) (b) $CaCl_2$ (1), $MgCl_2$ (2) and without salt (3) (c) $FeCl_3$ (1), $AlCl_3$ (2) and without salt (3) (d) $FeSO_4$ (1), $FeCl_3$ (2) and without salt (3) in presence of clay. (Values of FRET efficiencies were calculated from the spectra of figure 2 and figure 3).

**Table 1**

| Salt | FRET efficiency (E%) without clay | FRET efficiency (E%) with clay |
|---|---|---|
| KCl | 9.7% | 56.8% |
| $MgCl_2$ | 7.4% | 47.4% |
| $FeCl_3$ | 5.2% | 38.5% |

**Table 2**

| Salt | FRET efficiency (E%) with clay |
|---|---|
| KCl | 56.8% |
| NaCl | 62.15% |
| $MgCl_2$ | 47.4% |
| $CaCl_2$ | 41.1% |
| $FeCl_3$ | 38.5% |
| $AlCl_3$ | 45.6% |

**Table 3**

| Salt | FRET efficiency (E%) with clay |
|---|---|
| $FeSO_4$ | 46.8% |
| $FeCl_3$ | 38.5% |

**Table 4**

| Salts | FRET efficiency(E%) with clay | | |
|---|---|---|---|
| | salt con= 10 μM | salt con= 100 μM | salt con= 1000 μM |
| KCl | 56.8 | 48.7 | 41.2 |
| NaCl | 62.15 | 55.7 | 51.2 |
| $MgCl_2$ | 47.4 | 42.5 | 36.8 |
| $CaCl_2$ | 41.1 | 36.5 | 31.2 |
| $FeSO_4$ | 46.8 | 41.7 | 35.8 |
| $FeCl_3$ | 38.5 | 32.7 | 27.8 |
| $AlCl_3$ | 45.6 | 39.8 | 33.5 |

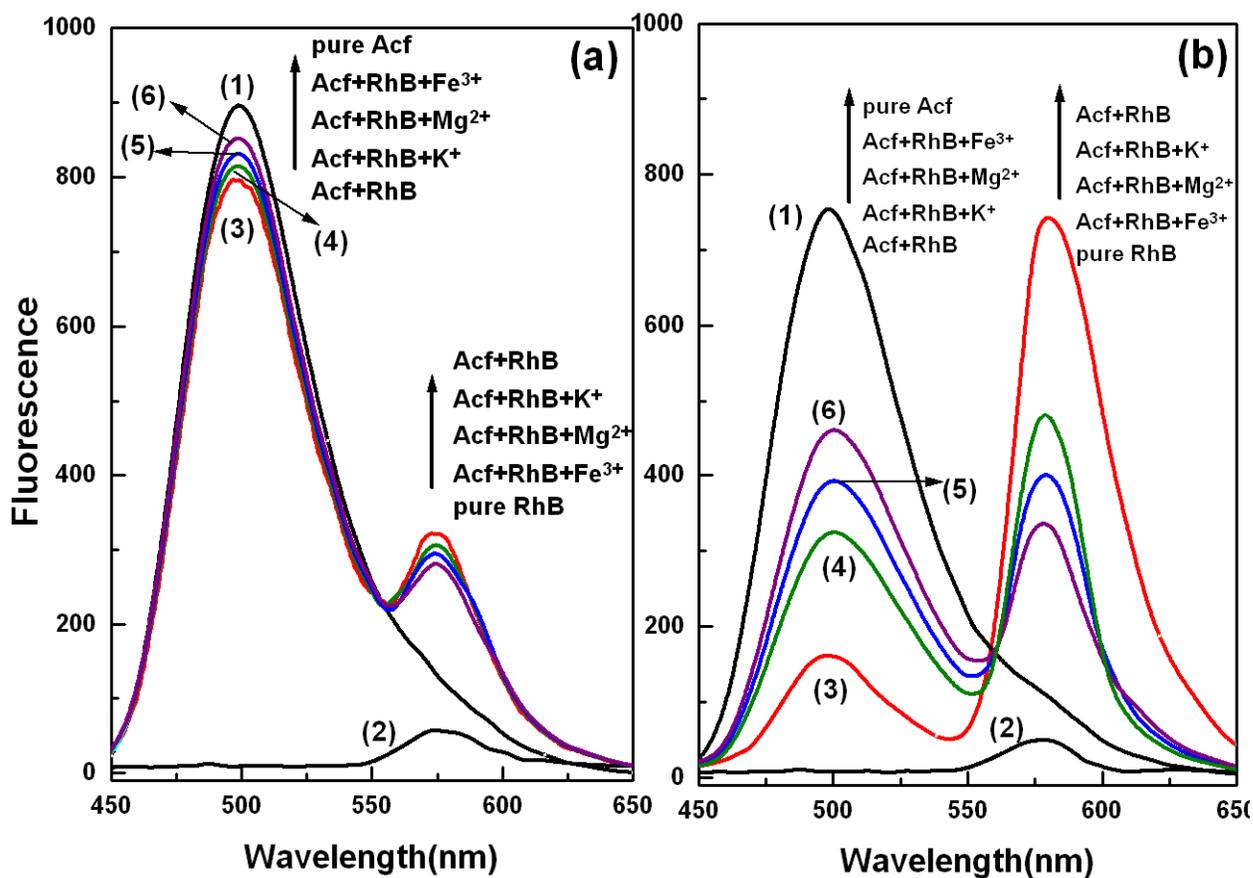

Fig. 1.

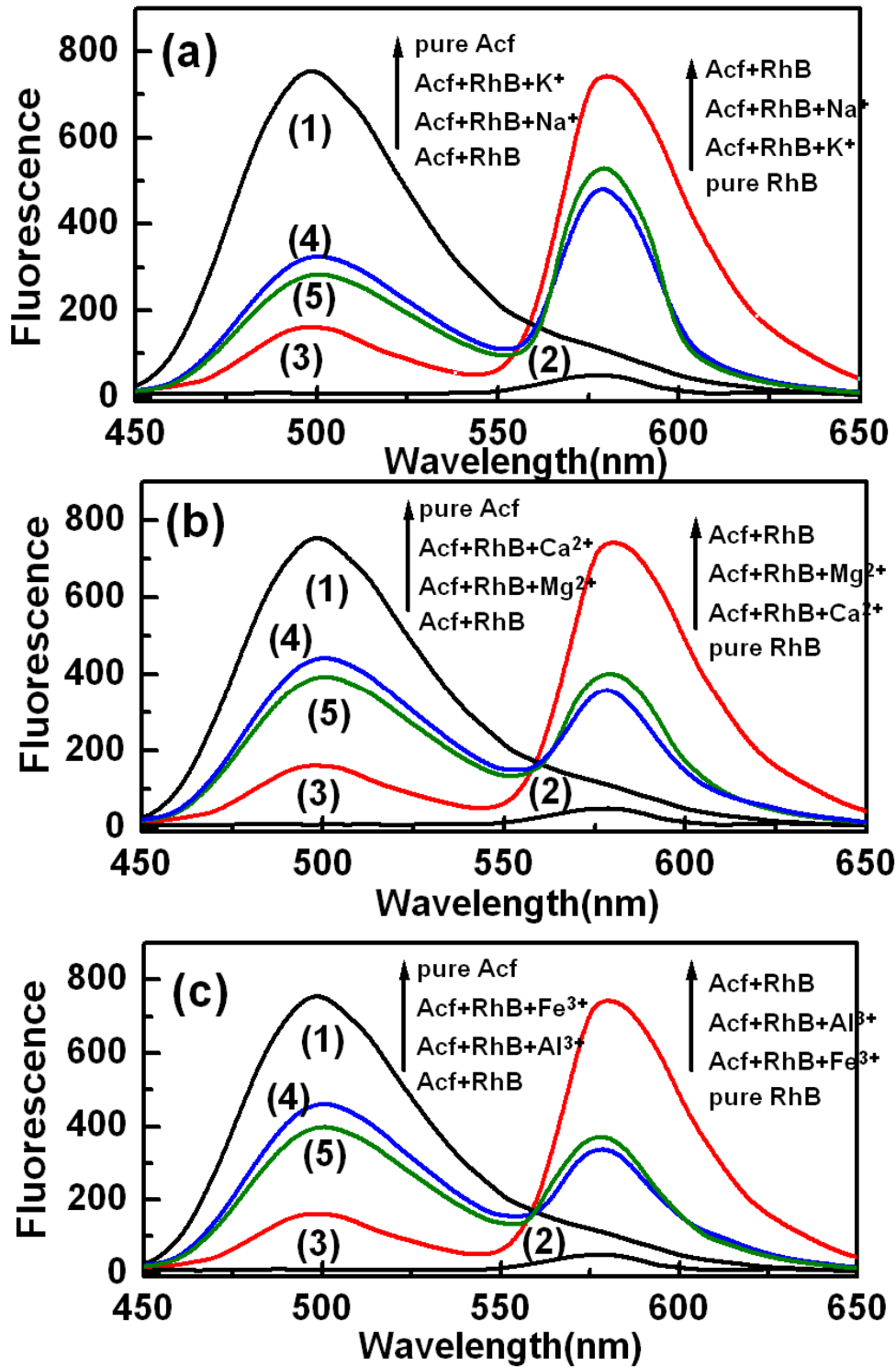

**Fig. 2.**

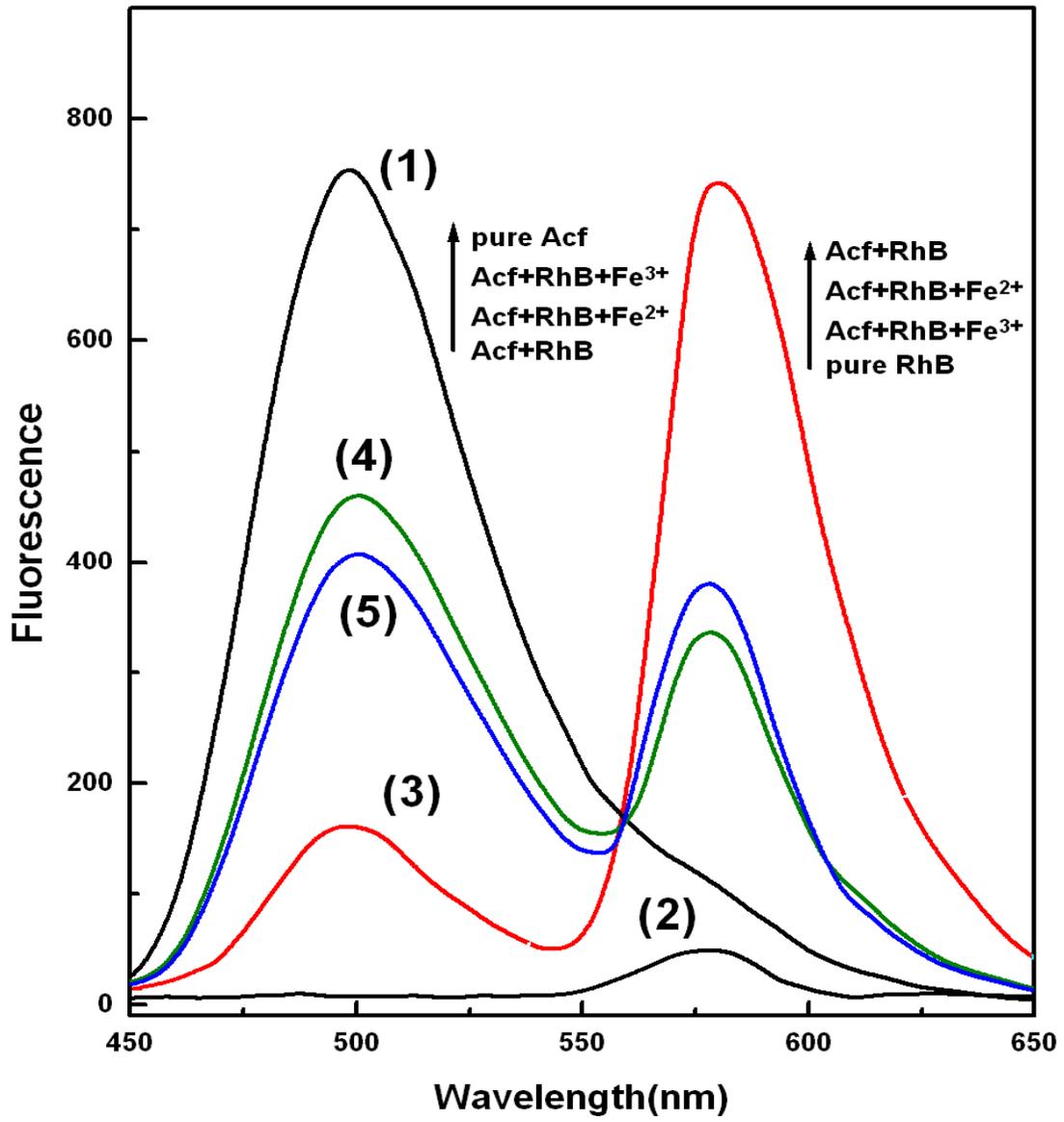

**Fig. 3.**

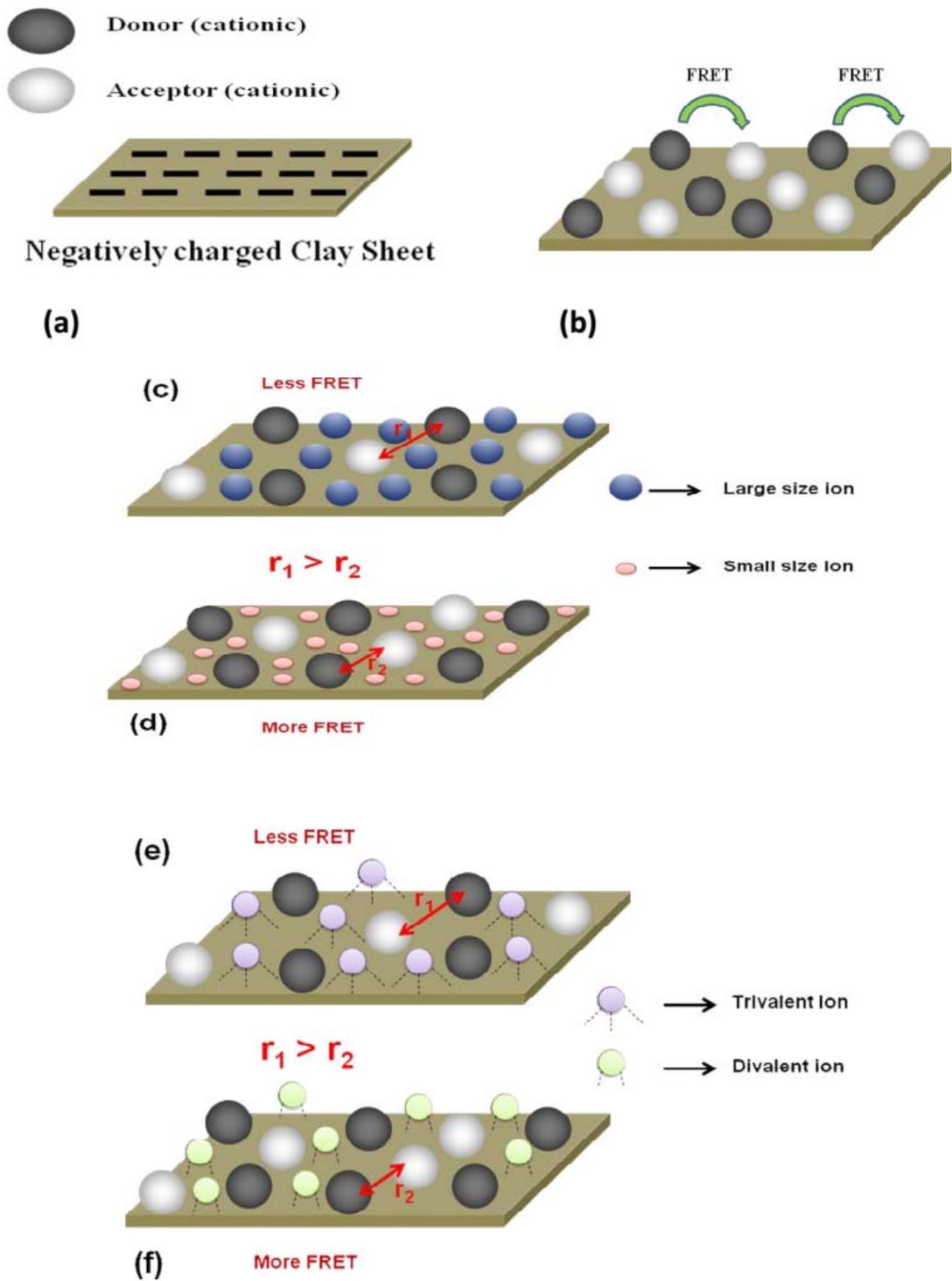

**Fig. 4.**

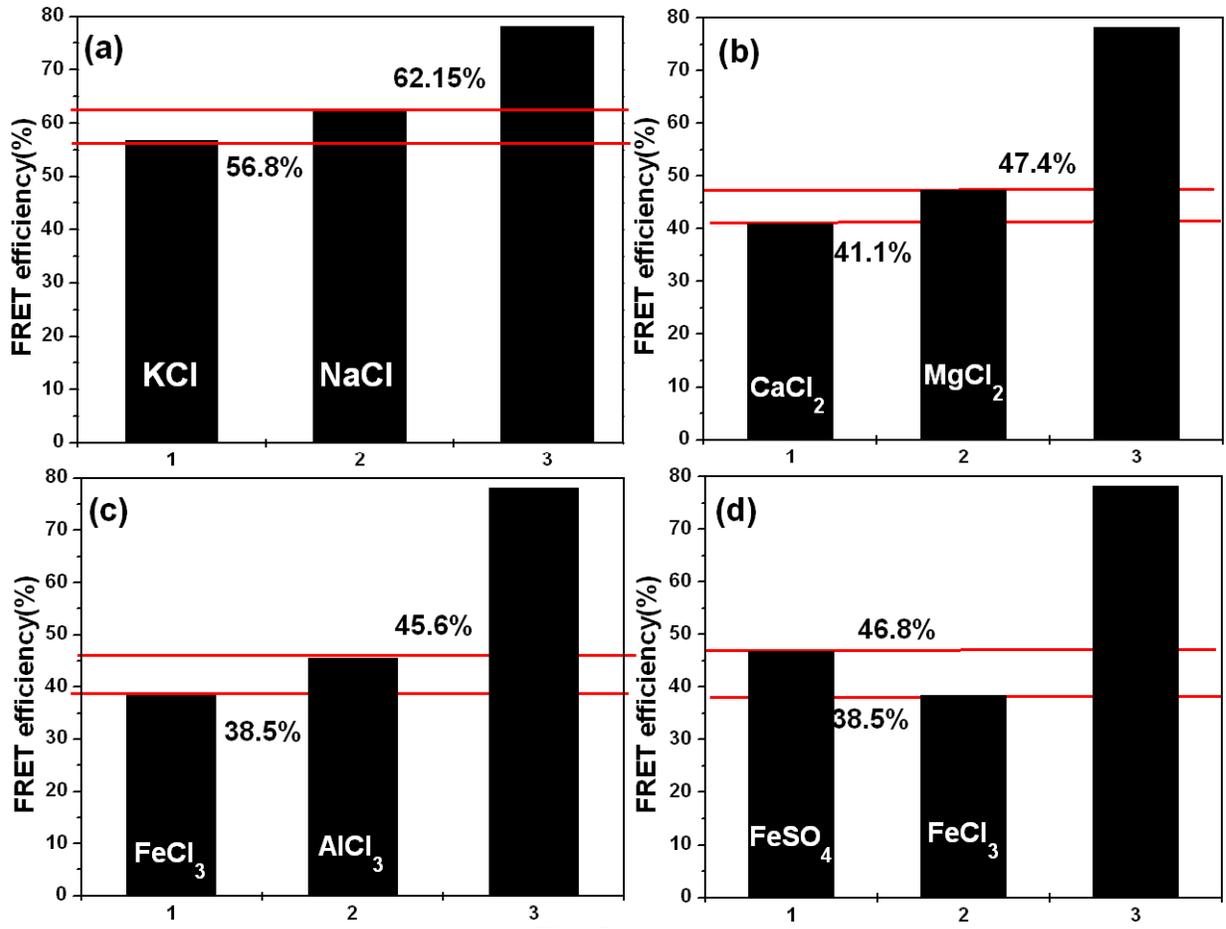

Fig. 5.